\documentclass[preprint,showpacs,superscriptaddress,preprintnumbers,amsmath,amssymb]{revtex4}

\usepackage{graphicx}
\usepackage{dcolumn}
\usepackage{amsmath}
\usepackage{amssymb}

\def\vk{{\bf k}}

\def\bra{\langle}
\def\ket{\rangle}

\newcommand{\eq}[1]{Eq.~(\ref{#1})}
\newcommand{\fig}[1]{Fig.~\ref{#1}}
\newcommand{\be}{\begin{equation}}
\newcommand{\ee}{\end{equation}}
\newcommand{\bea}{\begin{eqnarray}}

\newcommand{\eea}{\end{eqnarray}}
\newcommand{\bean}{\begin{eqnarray*}}
\newcommand{\eean}{\end{eqnarray*}}
\newcommand{\bfi}{\begin{figure}}
\newcommand{\efi}{\end{figure}}
\newcommand{\bc}{\begin{center}}
\newcommand{\ec}{\end{center}}
\newcommand{\ba}{\begin{array}}
\newcommand{\ea}{\end{array}}

\begin{document}


\title{Van Hove singularity and spontaneous Fermi surface symmetry breaking 
in Sr$_{3}$Ru$_{2}$O$_{7}$}

\author{Hiroyuki Yamase} 
\affiliation{Max-Planck-Institute for Solid State Research, 
Heisenbergstrasse 1, D-70569 Stuttgart, Germany}

\author{Andrey A. Katanin} 
\affiliation{Max-Planck-Institute for Solid State Research, 
Heisenbergstrasse 1, D-70569 Stuttgart, Germany}
\affiliation{Institute of Metal Physics, 620219 Ekaterinburg, Russia}



\date{\today}

\begin{abstract} 
The most salient features observed around a metamagnetic transition 
in  Sr$_{3}$Ru$_{2}$O$_{7}$ are well captured 
in  a simple model for spontaneous Fermi surface 
symmetry breaking under a magnetic field, 
without invoking a putative quantum critical point. 
The Fermi surface symmetry breaking happens in both a majority and a  
minority spin band but with a different magnitude of the order parameter, 
when either band is tuned close to van Hove filling by 
the magnetic field. 
The transition is second order for high temperature $T$ and 
changes into first order for low $T$. 
The first order transition is accompanied by a metamagnetic transition.  
The uniform magnetic susceptibility and 
the specific heat coefficient 
show strong $T$ dependence, especially a $\log T$ divergence at 
van Hove filling. 
The Fermi surface instability then  
cuts off such non-Fermi liquid behavior and 
gives rise to a cusp in the susceptibility and 
a specific heat jump at the transition temperature. 
\end{abstract}

\maketitle

Usually the symmetry of the Fermi surface (FS) 
satisfies the point-group symmetry of the underlying lattice structure. 
However, recently a symmetry-breaking Fermi surface deformation 
with a $d$-wave order parameter, where the FS expands along the $k_{x}$ 
direction and shrinks along the $k_{y}$ direction, or vice versa,  was 
discussed in the two typical models for cuprate superconductors, 
the $t$-$J$\cite{yamase00,edegger06} 
and Hubbard model\cite{metzner00,grote02} on a square lattice.
This $d$-wave type Fermi surface deformation ($d$FSD) is often called 
$d$-wave Pomeranchuk instability, referring to Pomeranchuk's stability 
criterion for isotropic Fermi liquids.\cite{pomeranchuk58} 
However, the $d$FSD often takes place even without violating such 
a criterion, since the $d$FSD instability is usually first order 
for low temperature,\cite{khavkine04,yamase05a} and it can also happen 
even for strongly correlated systems such as those described by 
the $t$-$J$ model.\cite{yamase00,edegger06}  
The $d$FSD instability is driven by forward scattering processes 
of electrons close to the van Hove points in the two-dimensional 
Brillouin zone. 
As a result, the symmetry of the electronic state is reduced from $C_{4v}$ 
to $C_{2v}$  while the lattice still retains $C_{4v}$ symmetry as long as 
no electron-phonon coupling is considered.  
The $d$FSD state has the same symmetry as the so-called electronic 
nematic order, which was widely discussed 
in the context of charge stripes for cuprates.\cite{kivelson03}

Recently such a novel FS instability was reported 
for the ruthenate compound 
Sr$_{3}$Ru$_{2}$O$_{7}$ in the presence of an external 
magnetic field.\cite{grigera04} 
Sr$_{3}$Ru$_{2}$O$_{7}$ is a bilayered material with metallic 
RuO$_{2}$ planes where the Ru ions form a square lattice. 
{\it Ab initio} calculations\cite{hase97} showed 
that the electronic band 
structure is similar to that for the single layered ruthenate 
Sr$_{2}$RuO$_{4}$, a well-known spin-triplet 
superconductor.\cite{mackenzie03}  
The ground state of Sr$_{3}$Ru$_{2}$O$_{7}$ is, however, 
paramagnetic.\cite{huang98} 
By applying a magnetic field $h$, 
Sr$_{3}$Ru$_{2}$O$_{7}$ shows a metamagnetic transition 
at $h=h_{c}$, around which non-Fermi liquid behavior is observed in 
various quantities:  resistivity,\cite{grigera01,perry01} 
specific heat,\cite{perry01,zxzhou04,perry05} 
and thermal expansion.\cite{gegenwart06} 
This non-Fermi liquid behavior was 
discussed in terms of a putative metamagnetic quantum critical 
end point (QCEP) and hence Sr$_{3}$Ru$_{2}$O$_{7}$ was often 
referred to as a system with a metamagnetic QCEP.\cite{millis02,grigera02,binz04} 
However, subsequent experiments\cite{grigera04} for ultrapure crystals 
showed that the hypothetical QCEP was hidden by a dome-shaped phase 
transition line; a first order transition 
was confirmed at the edges of the transition line and 
was accompanied by a metamagnetic transition, 
while a second order transition was inferred for high temperature $T$. 
Grigera {\it et al.}\cite{grigera04} associated this instability 
with a spontaneous $d$FSD. 
Later Kee and Kim\cite{kee05-misc} demonstrated 
a metamagnetic transition due to a first-order $d$FSD phase transition 
in a phenomenological model, which they discussed 
might be relevant to Sr$_{3}$Ru$_{2}$O$_{7}$. 
A direct experimental evidence of the $d$FSD, however, has not been 
obtained so far. 
Other scenarios, namely, microscopic phase separation due to 
Coulomb interaction\cite{honerkamp05} and magnetic domain formation 
due to long-range dipolar interactions,\cite{binz06} 
were also proposed to explain the experimental data for 
Sr$_{3}$Ru$_{2}$O$_{7}$.

In this Letter, 
in addition to confirming a metamagnetic transition 
due to a first-order $d$FSD phase transition as 
reported by Kee and Kim,\cite{kee05-misc}  
we show that the experimental phase diagram for Sr$_{3}$Ru$_{2}$O$_{7}$, 
and the $T$ dependences of the uniform magnetic susceptibility 
and the specific heat 
are well captured in terms of the 
$d$FSD instability near the van Hove filling, 
without invoking a putative QCEP. 
We analyze a simple model with a 
pure forward scattering interaction driving the $d$FSD instability 
in the presence of a magnetic field. 
We find that when either the majority or minority band is tuned to 
van Hove filling by the magnetic field, the $d$FSD instability occurs
in both bands, but with a different magnitude of the order parameter. 
The transition is second order for high $T$ and changes into 
first order for low $T$. The first order transition is accompanied by a 
metamagnetic transition. 
Both the magnetic susceptibility and the specific heat coefficient 
show strong $T$ dependences, 
especially a $\log T$ divergence at van Hove filling. 
This non-Fermi liquid behavior originates from the van Hove singularity 
in the density of states of the bare dispersion. 
The $d$FSD instability then cuts off such non-Fermi liquid behavior and 
produces a cusp in the susceptibility and a specific heat jump at the 
transition temperature.

We investigate the $d$FSD instability in the presence of a 
magnetic field on a square lattice. 
The minimal model reads 
\begin{equation}
 H = \sum_{\vk,\sigma} (\epsilon_{\vk}^{0}-\mu) \, n_{\vk}^{\sigma} + 
 \frac{1}{2N} \sum_{\vk,\sigma,\vk,\sigma'} f_{\vk\vk'} \, 
n_{\vk}^{\sigma} n_{\vk'}^{\sigma'} 
-h \sum_{\vk,\sigma}\sigma n_{\vk}^{\sigma} \, 
 \label{f+h-model}
\end{equation}
where $n_{\vk}^{\sigma} = c_{\vk\,\sigma}^{\dagger}c_{\vk\,\sigma}$ 
counts the electron number with momentum $\vk$ and spin $\sigma$; 
$c_{\vk\,\sigma}^{\dagger}$ ($c_{\vk\,\sigma}$) is an electron 
creation (annihilation) operator; $\mu$ is the chemical potential; 
$N$ is the number of lattice sites; $h$ is the magnetic field. 
For hopping amplitudes $t$ and  $t'$ between nearest and 
next-nearest neighbors on the square lattice, respectively, 
the bare dispersion relation is given by
\begin{equation}
 \epsilon_{\vk}^{0}= -2 t (\cos k_{x}+\cos k_{y}) 
 -4 t'\cos k_{x} \cos k_{y}\, . 
\end{equation}
The forward scattering interaction driving the spontaneous $d$FSD 
has the form 
\begin{equation}
 f_{\vk\vk'} =  - g \, d_{\vk} d_{\vk'} \,,   \label{fkk}
\end{equation}
with a coupling constant $g\geq 0$ and
a $d$-wave form factor $d_{\vk} = \cos k_x - \cos k_y$.
This ansatz mimics the structure of the effective interaction in 
the forward scattering channel as obtained for the 
$t$-$J$\cite{yamase00} and Hubbard model.\cite{metzner00} 
The model~(1) without the 
magnetic field was extensively studied in 
Refs.~\onlinecite{khavkine04}~and ~\onlinecite{yamase05a}.

We decouple the interaction by 
introducing a spin-dependent mean field  
$\eta^{\sigma}=-\frac{g}{N}\sum_{\vk} d_{\vk} \bra n_{\vk}^{\sigma} \ket$ 
and obtain a renormalized band dispersion  
$\xi_{\vk}^{\sigma}=\epsilon_{\vk}^{0}+\eta d_{\vk} - \mu^{\sigma}$ 
with $\eta=\sum_{\sigma} \eta^{\sigma}$; the mean fields are determined 
by minimizing the free energy, which is a even function with respect to 
$\eta$, and a solution of $\eta\geq 0$ is considered. 
The  $\sigma$-summed mean filed $\eta$ 
enters $\xi_{\vk}^{\sigma}$, and thus a finite $\eta^{\sigma}$ 
in general induces a finite $\eta^{-\sigma}$. 
The Zeeman field is absorbed completely in the effective 
chemical potential 
$\mu^{\sigma}=\mu+\sigma h$. 
Since our Hamiltonian~(\ref{f+h-model}) does not allow momentum transfer, 
the mean-field theory solves our model exactly in the 
thermodynamic limit. 

The magnetization is given by $m=\frac{1}{N}\sum_{\vk, \sigma}\sigma 
\bra n_{\vk}^{\sigma} \ket = \frac{1}{N}\sum_{\vk, \sigma}
\sigma f(\xi_{\vk}^{\sigma})$, and thus 
the uniform magnetic susceptibility $\chi$ is 
\be
\chi=\frac{\partial m}{\partial h} 
=- \frac{1}{N}\sum_{\vk,\sigma}f^{'}(\xi_{\vk}^{\sigma})\, ,
\label{chi}
\ee
where $f(x)=1/(e^{x/T}+1)$ is the Fermi function 
and $f'$ is its first derivative. 
The electronic specific heat coefficient 
$\gamma=C/T$ can be obtained straightforwardly by 
the second derivative of the 
free energy with respect to $T$ at fixed $\mu$,    
\bea
\gamma&=&-\frac{1}{T^{2}N}
\sum_{\vk,\sigma} (\xi_{\vk}^{\sigma})^{2} 
f'(\xi_{\vk}^{\sigma}) \nonumber \\
&&+ g\frac{\left(\frac{1}{TN}\sum_{\vk,\sigma} d_{\vk} \xi_{\vk}^{\sigma} 
f'(\xi_{\vk}^{\sigma}) \right)^{2}}{1+\frac{g}{N}\sum_{\vk,\sigma}d_{\vk}^{2}f'(\xi_{\vk}^{\sigma})}\,.
\label{gamma}
\eea
The second term \eq{gamma} is zero above $T_{c}$ and leads to a 
specific heat jump at $T_{c}$.

Band structure calculations for 
Sr$_{3}$Ru$_{2}$O$_{7}$\cite{hase97} 
yield 6 Fermi surfaces for $h=0$. We focus on a FS closest to 
$\vk=(\pi,0)$ and $(0,\pi)$ 
since the $d$FSD instability is driven by 
electrons near the van Hove points on a square lattice; 
we mimic such a FS by choosing $t'/t=0.35$. 
We fix $\mu/t=1$ and take $g/t=1$ for numerical convenience, although 
Sr$_{3}$Ru$_{2}$O$_{7}$ is expected to have a much smaller $g$ as 
we discuss later. 
We choose $t$ as a unit of energy so that 
$t=1$ in this paper. 
Since the results are symmetric with respect to $h\rightarrow -h$ and 
$\sigma \rightarrow -\sigma$, we consider only the case $h \geq 0$.  

\begin{figure}
\centerline{\includegraphics[width=0.4\textwidth]{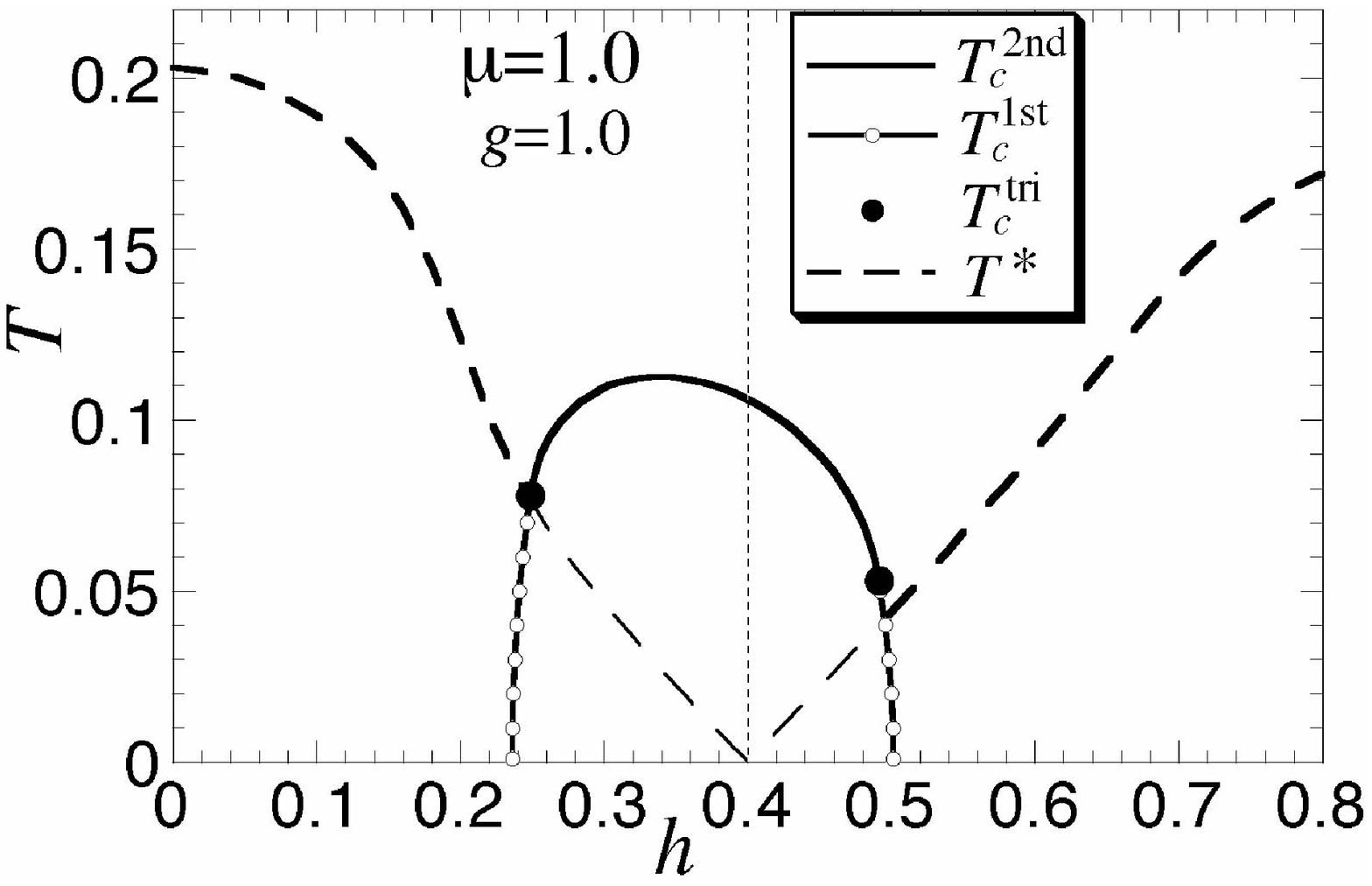}}
\caption{The $d$FSD phase diagram in the plane of magnetic field $h$ and 
temperature $T$ for $\mu=1.0$ and $g=1.0$; 
the transition is second order for high $T$ 
($T_{c}^{\rm 2nd})$ and first order for low $T$ ($T_{c}^{\rm 1st})$; 
end points of the second order line are tricritical points 
($T_{c}^{\rm tri})$; the dotted line ($h=0.4$) represents the van Hove 
energy of the up-spin band; the dashed line ($T^{*}$) 
denotes a peak position of the uniform magnetic susceptibility 
and the thin dashed line represents $T^{*}$ in the absence of the $d$FSD.} 
\label{phase-m1.0}
\end{figure}

\begin{figure}
\centerline{\includegraphics[width=0.4\textwidth]{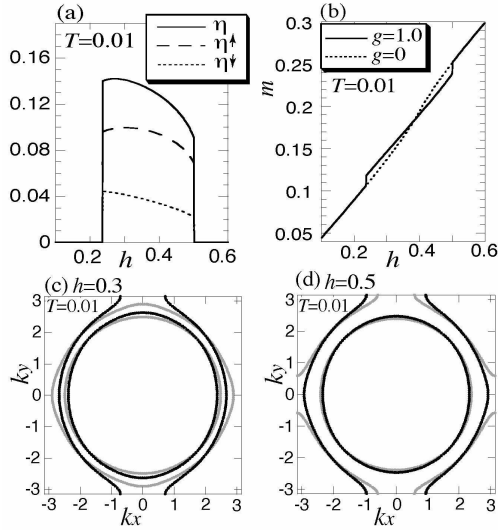}}
\caption{The mean-field solution at low $T$ for  $\mu=1.0$ and $g=1.0$. 
(a) $h$ dependence of the order parameter; 
note that $\eta=\eta^{\uparrow} + \eta^{\downarrow}$. 
(b) A metamagnetic transition due to the first order $d$FSD transition;  
the result for $g=0$ is also shown by a dotted line. 
FSs for $g=1$ (solid line) and $0$ (gray line) at 
$h=0.3$ (c) and $0.5$ (d); the deformation of the inner FS in (d) is 
hardly  visible.} 
\label{phase-m1.0-2}
\end{figure}

Figure~\ref{phase-m1.0} shows a phase diagram in the plane of applied 
magnetic field $h$ and temperature $T$. 
The $d$FSD transition occurs around the van Hove energy 
of the up-spin band ($h=0.4$) with 
a second order transition for high $T$ and a first order one for 
low $T$; end points of the second order line are 
tricritical points. 
Figure~\ref{phase-m1.0-2}(a) shows the 
$h$ dependence of the order parameter $\eta$, together with 
$\eta^{\sigma}$, at low $T$. 
Both $\eta^{\uparrow}$ and $\eta^{\downarrow}$ show a jump 
at the first order transition point, but with a different magnitude. 
The magnetization $m$ 
also shows a jump at the first order phase transition 
[\fig{phase-m1.0-2}(b)].
Its upward jump with increasing $h$ is due to a generic consequence 
of the concavity of the grand canonical potential 
as a function of $h$. Hence the first order transition of the $d$FSD 
instability is necessarily accompanied by a metamagnetic transition.  
The FSs at low $T$ are shown in Figs.~\ref{phase-m1.0-2}(c) and (d) 
for $h=0.3$ and $0.5$, respectively. 
The gray lines are FSs for $g=0$ and 
an outer (inner) FS corresponds to the up-spin (down-spin) band; 
the splitting of these FSs is due to the Zeeman energy. 
The FS instability drives a deformation of both FSs and 
typically leads to an open outer FS.

\begin{figure}
\centerline{\includegraphics[width=0.4\textwidth]{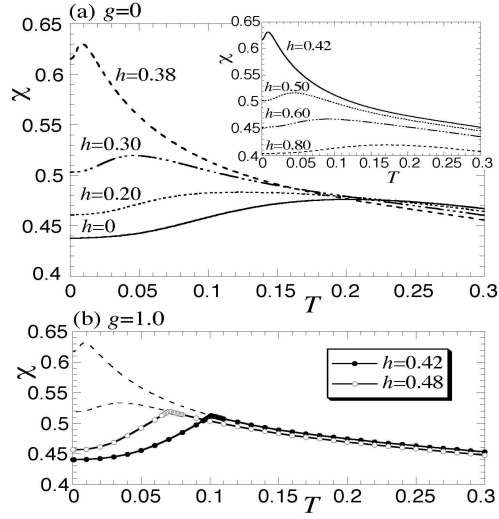}}
\caption{$T$ dependence of $\chi$ for several choices of $h$ for $g=0$ (a) 
and $g=1.0$ (b); the dashed lines in (b) are data for $g=0$; 
$h=0.4$ corresponds to the van Hove energy of 
the up-spin band.} 
\label{T-chi0}
\end{figure}

Figure~\ref{T-chi0} shows the $T$ dependence of $\chi$ [\eq{chi}] for 
several choices of $h$ for $g=0$. 
For a small field, $\chi$ has a weak $T$ dependence with a broad maximum 
at relatively high $T$ 
and becomes constant for low $T$, i.e. Pauli paramagnetic behavior. 
As $h$ moves close to the van Hove energy ($h=0.4$), 
$\chi$ starts to have a strong 
$T$ dependence and forms a pronounced peak at low temperature. 
The peak position is pushed down to zero temperature 
at the van Hove energy, where a $\log T$ divergence appears. 
Similar behavior is seen when the magnetic field 
is reduced from a large $h$ to the 
van Hove energy [inset of \fig{T-chi0}(a)]. 
Defining $T^{*}$ as the peak position of $\chi$, we thus obtain 
the V-shaped $T^{*}$ line shown in \fig{phase-m1.0}; 
in particular $T^{*}$ goes to zero at the van Hove energy. 
This behavior is due to the van Hove singularity of the up-spin band, 
not due to an underlying quantum critical point. 
Around van Hove filling, 
the $d$FSD instability occurs and produces a cusp in 
the $T$ dependence of $\chi$ as shown in \fig{T-chi0}(b). 
A $T^{*}$ line is thus not defined inside the symmetry broken phase 
and the thin dashed line in \fig{phase-m1.0} represents $T^{*}$ in the 
absence of the $d$FSD.

Figure~\ref{phase-m1.0} is very similar to the 
phase diagram reported by Grigera {\it et al.}\cite{grigera04} with 
a first order transition 
for low $T$, a second order transition for high $T$, and  
a V-shaped $T^{*}$ line, which crosses the transition line 
near the tricritical points. 
In experiments, the $T^{*}$ line was determined by 
thermal expansion measurements.\cite{mackenzie-misc} 
While peak positions of the magnetic susceptibility were not 
studied systematically in experiments,\cite{mackenzie-misc} 
we expect that the magnetic susceptibility shows 
a similar V-shaped $T^{*}$ line around the van Hove energy. 
On the other hand, 
a strong $T$ dependence of the magnetic susceptibility [\fig{T-chi0}(a)] 
was observed in experiments;\cite{ikeda00}  
the cusp of $\chi$ shown in \fig{T-chi0}(b) can be tested. 
Experimental energy scales are, 
however, much smaller than in our result. 
To obtain a comparable $T_{c} \sim 1{\rm K}$, the coupling constant $g$ 
should be reduced substantially. 
As clarified in Ref.~\onlinecite{yamase05a}, 
in the weak coupling limit the dome-shaped transition line of the $d$FSD 
is  
characterized by a single energy scale 
$\epsilon_{\Lambda}e^{-1/(2\overline{g})}$,  
where $\epsilon_{\Lambda}$ 
is a cutoff energy and 
$\overline{g}=2m^{*}g/\pi^{2}$ is a dimensionless coupling constant 
with the effective mass $m^{*}$ near the van Hove energy. 
The result in the weak coupling limit was checked to be applicable 
even for finite $g$ with good accuracy.\cite{yamase05a} 
Therefore the phase diagram in \fig{phase-m1.0} does not change 
essentially for a smaller $g$, which just reduces 
the energy scale of the $d$FSD transition line 
around the van Hove point. 
The relative position of $T^{*}$ and 
$T_{c}^{\rm tri}$ also does not change appreciably, since $T^{*}$ 
has linear  dependence of $h$ around the van Hove energy. 

\begin{figure}
\centerline{\includegraphics[width=0.4\textwidth]{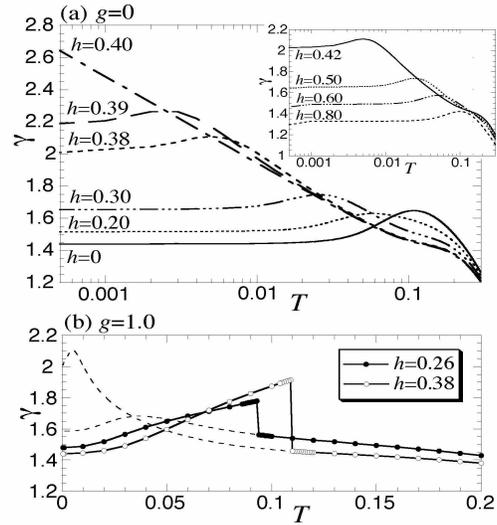}}
\caption{$T$ dependence of $\gamma$ for several choices of $h$ 
for $g=0$ (a) and $g=1.0$ (b); a logarithmic $T$ scale is used in (a); 
the dashed lines in (b) are data for $g=0$.}  
\label{T-gamma}
\end{figure}

The electronic specific heat 
is calculated exactly in our model [see \eq{gamma}].  
We first show $\gamma$ for $g=0$ in \fig{T-gamma} 
as a function of $T$ for several choices of $h$. 
For a relatively small field, 
$\gamma$ shows a peak structure for high $T$ 
(in a logarithmic scale), but becomes constant for 
lower $T$ as expected for the normal Fermi liquid. 
As $h$ moves close to the van Hove energy ($h=0.4$),  
the peak position is shifted to lower $T$ and 
a $\log T$ divergence appears at the van Hove energy. 
A similar behavior is also seen when $h$ is reduced from 
a large value to the van Hove energy [inset in \fig{T-gamma}(a)]. 
For a finite $g$, the $d$FSD instability 
produces a jump in the specific heat at the transition temperature 
as shown in \fig{T-gamma}(b).  
This jump typically becomes larger as $h$ is closer to the van Hove energy. 
Unlike the situation in the BCS theory, 
the ratio of the magnitude of the jump $\Delta\gamma$ 
and the normal state specific heat $\gamma_{n}$ at $T_c$ is not a 
universal value. In particular, in the weak coupling limit 
$T_{c}$ scales as $\epsilon_{\Lambda}e^{-1/(2\overline{g})}$  
as discussed above.  
Since $\gamma$ shows a $\log T$ dependence at van Hove filling, 
we obtain $\gamma_{n} \sim \log T_{c} \sim \overline{g}^{-1} 
\propto g^{-1}$, while $\Delta\gamma$ is a certain finite value.

A $T$ dependence of $\gamma$ 
very similar to \fig{T-gamma}(a) was actually obtained 
in experiments.\cite{perry01,zxzhou04,perry05}  
The specific heat jump is not observed in experiments, but 
may require 
a more precise measurement, which 
would provide a definite evidence of the second order transition; 
the authors in Ref.~\onlinecite{grigera04} inferred a  
second order transition from the magnetization and the resistivity 
measurements.

We have shown that the most salient features observed 
in Sr$_{3}$Ru$_{2}$O$_{7}$ are 
well captured in terms of the $d$FSD instability near van Hove filling. 
In particular, the non-Fermi liquid behavior reported for the 
$T$ dependence of $\chi$ and $\gamma$ can be associated with 
the van Hove singularity, not with a putative QCEP as usually 
discussed.\cite{millis02,grigera02}  
Around van Hove filling, it is known\cite{honerkamp02,kampf03} that 
various other ordering tendencies develop  
and probably compete with the $d$FSD tendency. 
We may thus allow other interactions such as
ferromagnetism and antiferromagnetism in our model 
to explore more detailed comparison with experimental data as well as 
the interplay among various ordering tendencies 
in the presence of the magnetic field. 
The most interesting future issue is how the anomalous $T$ dependence 
of the resistivity\cite{grigera01,perry01} 
observed in Sr$_{3}$Ru$_{2}$O$_{7}$ above $T_{c}$ can be understood. 
A crucial question is whether classical $d$FSD fluctuations 
and the van Hove singularity are sufficient to capture 
the resistivity data or whether quantum critical fluctuations 
originating from some QCEP are necessary. 
In the latter case, 
how is the QCEP related to the $d$FSD instability and the van Hove 
singularity that we have discussed in the present paper? 
Both the chemical potential and the magnetic field 
have to be fine-tuned to reach a QCEP while it is sufficient for either 
of them to be tuned to realize van Hove filling.  
In this sense, a QCEP scenario imposes an additional 
constraint on Sr$_{3}$Ru$_{2}$O$_{7}$.

When this work was complete, we have learned about a 
recent experimental observation of a large magnetoresistive 
anisotropy inside the dome-shaped transition line 
in Sr$_{3}$Ru$_{2}$O$_{7}$,\cite{borzi06} 
consistent with the existence of Fermi surface symmetry breaking 
in this compound.

We are grateful to 
C. Honerkamp, A. P. Mackenzie, and R. S. Perry  
for valuable discussions, and 
especially to W. Metzner for a critical reading of the manuscript. 


\bibliography{main.bib}

\newpage

\end{document}